\documentclass{article}
\usepackage{spconf,amsmath,amsfonts,bm,graphicx}
\usepackage{multirow,makecell}

\usepackage{pgfplots}
\usepackage{tikz}
\usepackage{wrapfig}
\usetikzlibrary{matrix}
\pgfplotsset{compat=1.14}

\definecolor{bblue}{HTML}{4F81BD}
\definecolor{rred}{HTML}{C0504D}
\definecolor{ggreen}{HTML}{9BBB59}
\definecolor{ppurple}{HTML}{9F4C7C}

\def\eg{\textit{e.g.\ }}
\def\ie{\textit{i.e.\ }}

\title{A LIGHT-WEIGHT MULTIMODAL FRAMEWORK FOR IMPROVED ENVIRONMENTAL AUDIO TAGGING}
\name{Juncheng Li$^{\dag\ddag}$, 
Yun Wang$^{\ddag}$,
Joseph Szurley$^{\dag}$,
Florian Metze$^{\ddag}$,
Samarjit Das$^{\dag}$
\thanks{This work is primarily supported by the Bosch Graduate Research Fellowship to the School of Computer Science at CMU.
This work used the ``bridges'' cluster (at PSC) of the
Extreme Science and Engineering Discovery Environment (XSEDE) \cite{XSEDE},
supported by NSF grant number ACI-1548562.}}
\address{\{junchenl,yunwang,fmetze\}@cs.cmu.edu, \{joseph.szurley,samarjit.das\}@us.bosch.com\\
$^{\dag}$ Robert Bosch LLC, Research and Technology Center, USA  \\
$^{\ddag}$Language Technology Institute, Carnegie Mellon University, Pittsburgh, USA}

\begin{document}
\ninept

\maketitle

\begin{abstract}

The lack of strong labels has severely limited the state-of-the-art fully supervised audio tagging systems to be scaled to larger dataset.
Meanwhile, audio-visual learning models based on unlabeled videos
have been successfully applied to audio tagging,
but they are inevitably resource hungry and require a long time to train.

In this work, we propose a light-weight, multimodal framework for environmental audio tagging.
The audio branch of the framework is a convolutional and recurrent neural network (CRNN)
based on multiple instance learning (MIL). It is trained with the audio tracks of a large collection of weakly labeled YouTube video excerpts;
the video branch uses pretrained state-of-the-art image recognition networks
and word embeddings to extract information from the video track
and to map visual objects to sound events.

Experiments on the audio tagging task of the DCASE 2017 challenge
show that the incorporation of video information
improves a strong baseline audio tagging system by
5.3\% in terms of $F_1$ score.
The entire system can be trained within 6~hours on a single GPU,
and can be easily carried over to other audio tasks such as speech sentimental analysis.
\end{abstract}


\section{Introduction}
\label{sec:intro}

Environmental audio tagging is the task of labeling audio recordings with the types of sound events occurring in them. Most conventional audio tagging systems (\eg~\cite{li2017comparison}) use supervised learning models that require strong labels, which specify the time span of each event. However, strongly labeled datasets are scarce and usually domain specific, which not only limits the potential for supervised systems trained on these datasets to be scaled up, and also severely impacts their generalizability to other domains.

To tackle this problem, the audio research community have turned to large-scale dataset with weak labels. Compared with strongly labeled datasets, weakly labeled datasets are much less expensive to collect at scale and can cover a wider range of sound event types.  In the weakly labeled datasets, the time spans of the events are not explicitly given, and only the presence or absence of events is known.  For example, Google Audio Set~\cite{gemmeke2017audio} encompasses a variety of human and animal sounds, musical instruments and genres, and common everyday environmental sounds. Systems developed on these datasets can be suitable for a larger number of domains, such as speech sentimental analysis in paralinguistics.

Without knowing the onset and offset times of the sound events, it is difficult to learn the acoustic characteristics of sounds because we do not know which parts of a clip reflect its label.
Multiple instance learning (MIL)~\cite{amores2013multiple} provides a way to link the clip-level labels with the individual frames of the clips, and thus enables supervised learning on the frame level. In MIL, each clip is regarded as a bag, and the frames as instances in the bag. A bag is positive if any instance in it is positive, and negative if all the instances in it are negative. 

In the mean time, the vision research community are trying to explore the correlation between audio and video. The field of computer vision has been revolutionized by the emergence of massive labeled datasets~\cite{russakovsky2015imagenet} and learning deep representations~\cite{krizhevsky2012imagenet,simonyan2014very}. Recent progress in this field has enabled machines to recognize scenes and objects in images and videos accurately, and a number of pre-trained models with very good accuracy~\cite{simonyan2014very,szegedy2017inception} have been made available to the public.
The success in visual recognition was well-translated into the recognition of audio: \cite{aytar2016soundnet} achieved a significant performance improvement from the state-of-the-art of audio-tagging models by transferring discriminative visual knowledge from well established visual recognition models into the sound modality using unlabeled video as a bridge; \cite{ArandjelovicZ17} trained visual and audio networks from matching audio and video pairs without any supervision, and the networks exhibited performance even superior to supervised models on audio tagging tasks; \cite{owens2016visually} showed that it was possible to synthesize audio tracks for silent videos of objects interacting with materials that could deceive the human ear. However, all of these methods require long hours and significant computation resources to train.

We consider this question: What is the best of both worlds? Can we leverage knowledge from both the audio and vision communities to help us with audio tagging?
In this work, we propose a novel multimodal framework that incorporates information from both the audio and video modalities without incurring excessive computational cost. On one hand, we train a CRNN network using the audio data to produce clip-level predictions based on the multiple instance learning (MIL) paradigm.
On the other hand, we extract key frames from the video data to summarize video clips, apply the pre-trained Inception V4 model~\cite{szegedy2017inception} to recognize the visual objects in the key frames, and then use pretrained GloVe vectors~\cite{pennington2014glove} to map the video labels to sound events.
The final predictions are made by fusing the knowledge in the two branches.
We evaluate our multimodal system on the audio tagging task of the DCASE 2017 challenge~\cite{Badlani2017}, and show that our proposed framework improves over a baseline audio-only system by 5.3\% in terms of $F_1$ score. In addition, our framework is light-weight and can be trained within 6 hours on a single GPU.

\section{Methodology}
\label{sec:method}

\subsection{Multiple Instance Learning for Audio Tagging}

\begin{figure}[t]
\centering
\includegraphics[width=\linewidth]{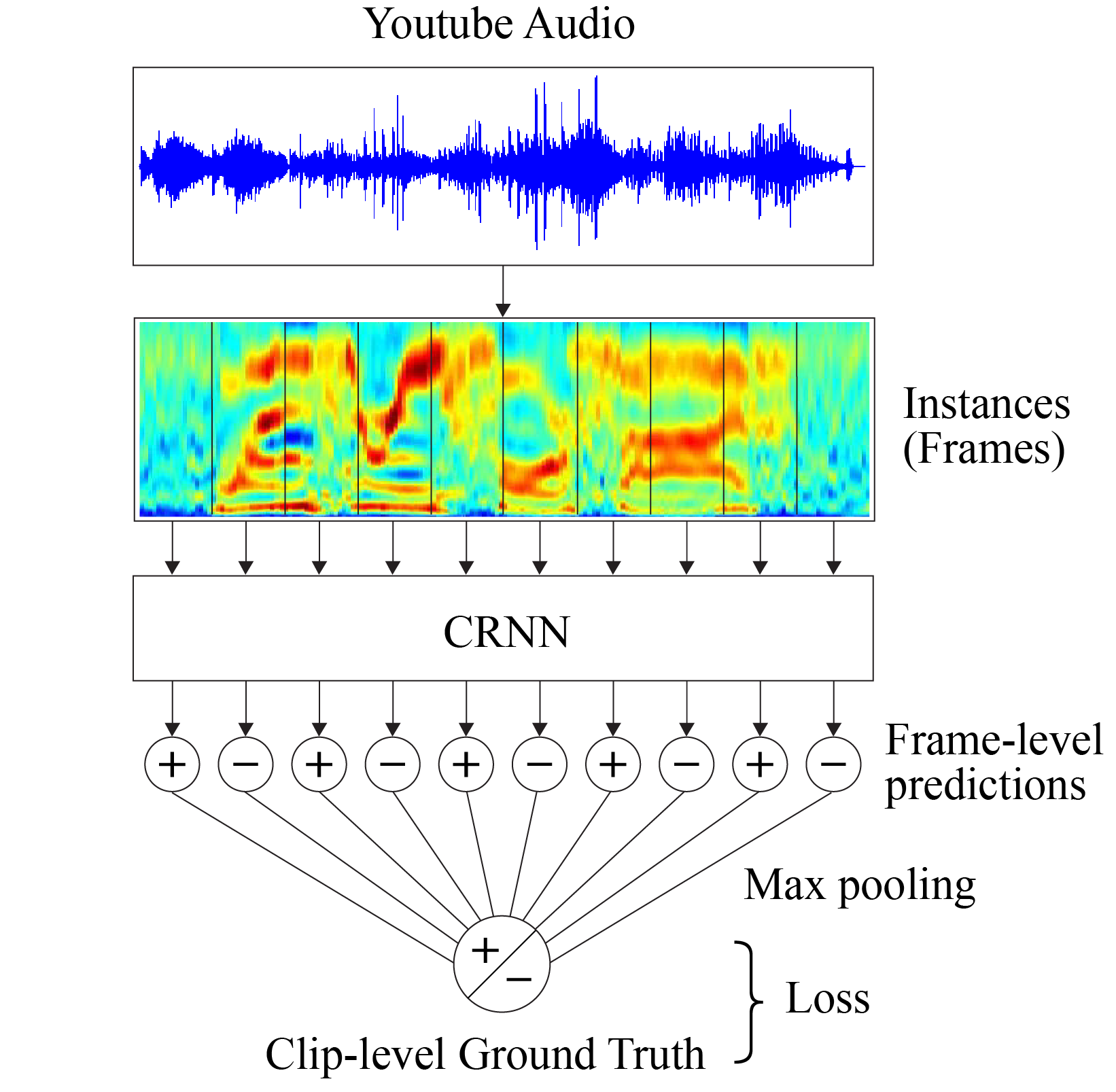}
\caption{Block diagram of the MIL system for audio tagging. Note here, each frame only has positive and negative prediction.}
\label{fig:MIL}
\end{figure}

Audio tagging based on weakly labeled data can be formulated as a multiple instance learning (MIL) problem. In MIL, the instance-level label is not known; instead, the instances are grouped into \emph{bags}, and we only have labels for the bags. The relationship between the label of a bag and the (unknown) labels of the instances in it follows the \emph{standard multiple instance assumption}: a bag is positive if it contains at least one positive instance, and is negative if all the instances are negative. In the audio clip tagging task, we can treat each clip as a bag, and the audio frames of the clip as instances. We predict the probability of each event being active in each frame; for each event, we aggregate the frame-level predictions using a max pooling function to obtain the probability of the event being active in the clip:
\[
    \hat{P}(e_k|\mathbf{x}_i) = \max_j \hat{P}(e_k|x_{ij})
\] 
where $\hat{P}(e_k|\mathbf{x}_i)$ is the predicted probability of event $k$ in the entire $i$-th clip, and $\hat{P}(e_k|x_{ij})$ is the predicted probability of event $k$ in the $j$-th frame of the $i$-th clip. 
The loss function can be constructed by comparing the clip-level prediction for each event with the clip-level label $y_{ik}$, which is 1 if the event is present and 0 otherwise. We employ a cross entropy loss averaged over all clips and all event types, which is
\begin{align*}
    L(\mathbf{x}) = -\frac{1}{NK} \sum_{i,k} \{& y_{ik} \log \hat{P}(e_k|\mathbf{x}_i) + \\
    & (1 - y_{ik}) \log [1 - \hat{P}(e_k|\mathbf{x}_i)] \}
\end{align*}
The loss function can be minimized with any gradient-based algorithm, but during back-propagation, only the maximally scoring frame receives a gradient and passes it down to the frame-level classifier. The structure of the MIL system using max pooling~\cite{amores2013multiple} is shown in Fig.~\ref{fig:MIL}.

\subsection{Learning Sounds from Visual Information}
\label{sec:visual}

Because the audio and video tracks of video clips are
often correlated and synchronized,
we can also extract information from the video track
that are indicative of the sounds in the audio track.
We try to predict sound events from the video track
using the following steps,
which are shown in the right branch of Fig.~\ref{fig:rerank}.

\subsubsection{Key Frame Extraction}

The video track of each clip contains many video frames,
some of which are nearly identical while others are distinct.
Selecting a subset of frames at equal intervals can
reduce the burden of subsequent object recognition,
but the frames selected this way are not guaranteed to
well represent the entire clip.
In~\cite{mithun2017trecvid}, it has been found important to
select a representative set of frames (called \emph{key frames})
to preserve the information in the video track.
This is a clustering problem and could be solved using
conventional clustering algorithms such as k-Means;
we used the following sparse coding algorithm~\cite{elhamifar2016dissimilarity}
to select a representative set because it is more robust to outliers compared with K-Means.

Given a set of objects $X = \{x_1, \ldots, x_n\}$,
the sparse coding algorithm operates on a dissimilarity matrix $D \in \mathbb{R}^{n \times n}$,
where the entry $d_{ij}$ measures the dissimilarity between the objects $x_i$ and $x_j$.
In order to express the relationship of ``representing'' between the objects,
a binary membership matrix $Z \in \mathbb{R}^{n \times n}$ is defined,
in which $z_{ij} = 1$ means that $x_i$ is a representative for $x_j$.
The algorithm aims to minimize the total dissimilarity between each object
and its representative, \ie $\sum_{i,j} z_{ij} d_{ij}$.
To ensure that each object is represented by one and only one object,
the membership matrix $Z$ must have each column normalized, \ie $\sum_i z_{ij} = 1, \forall j$;
to restrict the number of representative objects,
the number of non-zero rows of $Z$ must be as few as possible.
The sparse coding algorithm therefore tries to solve the following optimization problem:
\[
\min_{\{z_{ij}\}} \lambda \sum_{i=1}^n I(z_i) + \sum_{i=1}^n \sum_{j=1}^n z_{ij} d_{ij}
\]
\[
\text{s.t.} \sum_{i=1}^n z_{ij} = 1, \forall j; z_{ij} \in \{0,1\}, \forall i,j
\]
where $I(z_i) = 0$ if the $i$-th row of $Z$ is all zeros and $1$ otherwise,
and $\lambda$ is a regularization parameter that controls the size of the representative set.
Because the optimization is NP hard when the $z_{ij}$ must be binary,
the algorithm actually solves a relaxed version of the problem in which
$z_{ij}$ can be any real number in $[0,1]$.
The representatives can be selected by summing up each row of the matrix $Z$,
and picking the top few rows with the largest sums.

In order to select key frames quickly, it is necessary
to find an efficient way of encoding the frames into feature vectors.
We use the light-weight, convolutional AlexNet~\cite{krizhevsky2012imagenet} for this purpose:
we feed each video frame into the network and extract features from its ``conv5'' layer.
The dissimilarity matrix is made up of the Euclidean distances between the feature vectors.
We select 4 key frames for each clip.

\subsubsection{Object Recognition and Knowledge Mapping}

Once we get the key frames, we can pass them through the pretrained
InceptionNet V4 model~\cite{szegedy2017inception}
to recognize the objects in them.
For each key frame, this produces a distribution over 1,000 classes.
To reduce the noise in the probabilities of unlikely classes,
we ``rectify'' these distributions by retaining only
the top 10 classes and renormalizing their probability mass among themselves.
To obtain a clip-level representation of the video information,
we sum up the rectified distributions of all the key frames,
and rectify the result again.
This yields 10 object classes and their probabilities.

Now we have a belief of what objects are present in the video track of a clip,
we want to map it to the confidence of the sound events we are interested in.
We conduct this knowledge mapping using pre-trained GloVe vectors (glove.840B.300d)~\cite{pennington2014glove}. Since pre-trained GloVe vectors already contain a vector for each vocabulary from its knowledge source, we adopt the GloVe vectors from the descriptions of all the
visual objects and sound events, and compute the cosine similarity of each (object, sound event) pair.
For each of the top 10 object classes,
we assign its probability to the sound event
that is closest to it in terms of cosine similarity;
some sound events may receive probabilities from more than one object class.
This results in an estimation of the probability of each sound event
in the clip as indicated by the video track.

\subsection{Fusion of Audio and Visual Knowledge}

Both the audio track and the video track predict
a probability for each sound event.
The final fusion step tries to produce a refined prediction
by combining the two knowledge sources.
This combination is implemented with a weighted average
of the two probabilities, where the weights are tunable for each sound event. The better performing model on each individual class would have a higher weight on that class, and the weigtht combination is tuned toward achieving the best overall $F_1$ score.

\begin{figure}[t]
\includegraphics[width=\linewidth]{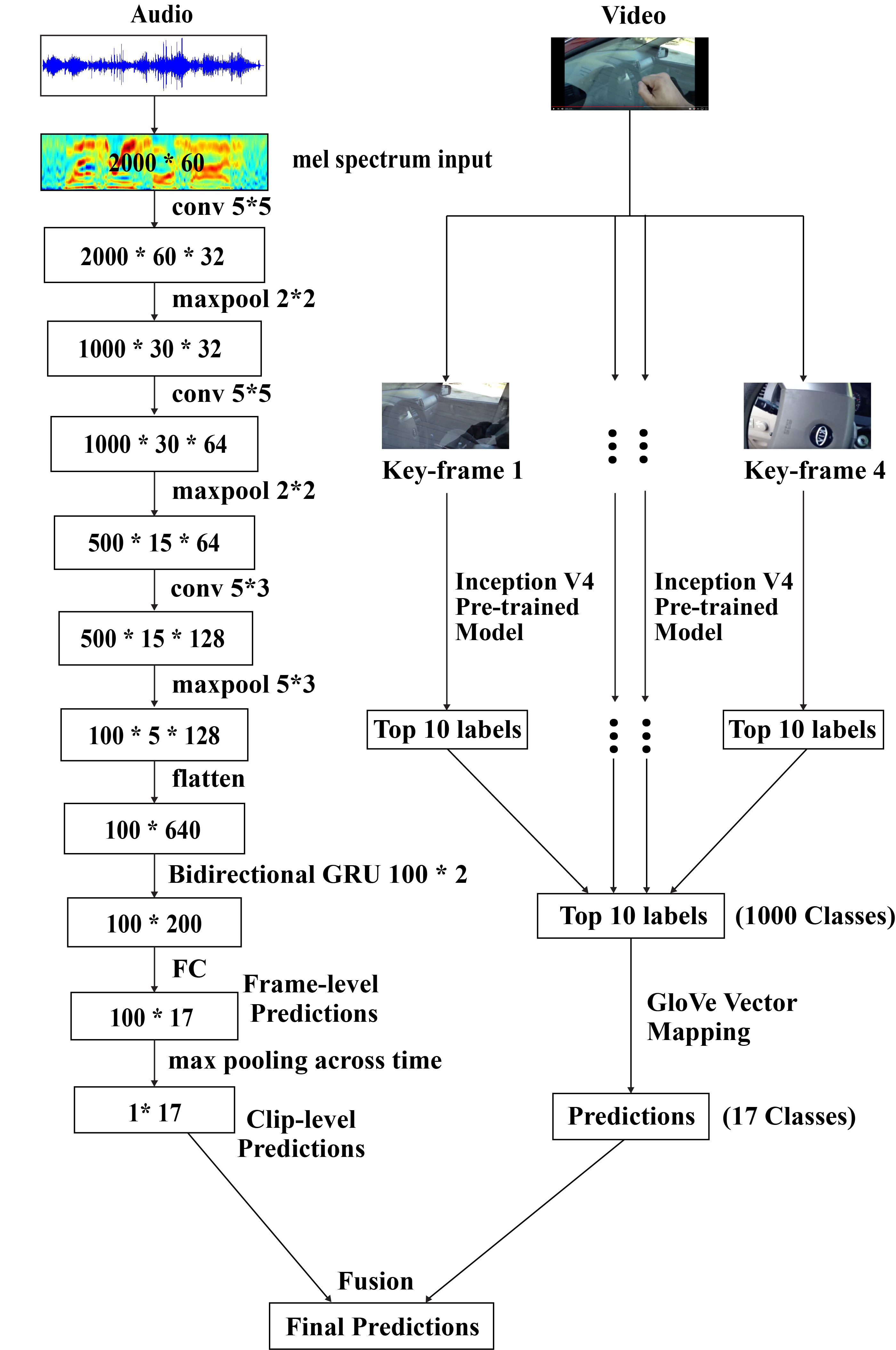}
\caption{Block diagram of our proposed multimodel framework. Note here that both Audio and Video bottleneck features contain dynamics.}
\label{fig:rerank}
\end{figure}

\section{Experiment}
\label{sec:exp}

\subsection{Dataset}

We evaluated our framework on the data for Task~4 of the DCASE 2017 challenge~\cite{Badlani2017}.
This data is a subset of the Google Audio Set~\cite{gemmeke2017audio},
which is a collection of over 2~million 10-second YouTube video clips,
annotated with the types of sound events occurring in them.
The annotation covers more than 600 audio event types,
but it is weak in the sense that no information about the time span of the events is known.

The DCASE 2017 data focuses on a subset of 17 sound event types;
these include vehicle and warning sounds that are interesting for self-driving cars and smart cities.
The corpus consists of a training set of 51,172 clips (around 142 hours),
a public test set of 488 clips, and a private evaluation set of 1,103 clips.
The test and evaluation sets are balanced across the 17 sound event types,
while the training set is not.
We reserved 1,142 clips from the training set to make a balanced validation set,
which we used for hyperparameter tuning.

\subsection{The Baseline Audio Tagging System}

The structure of the baseline audio tagging system, which only uses the audio track,
is shown in the left branch of Fig.~\ref{fig:rerank}.

The input to the system is Mel spectral features.
We first resampled the raw waveform to 20~kHz,
and then divided it into frames of 24~ms (480~samples)
with a frame shift of 5~ms.
We then applied a 1,024-point Fourier transform to each frame,
and aggregate the output into a 60-dimensional Mel spectral feature vector.
In this way, each 10-second clip is represented by a 2000 $\times$ 60 feature matrix.

The feature matrix is treated as a two-dimensional image,
and passed through a series of convolutional and local max pooling layers.
After the pooling layers, the feature map sequence represent an equivalent frame rate down to 10Hz from original raw frame rate of 200Hz.
The output of the last pooling layer is fed into a bidirectional Gated Recurrent Unit (GRU) layer
with 100~neurons in each direction.
Dropout with a rate of 0.2 was applied after each pooling layer and the GRU layer.
Finally, a fully connected layer with 17~neurons and the sigmoid activation function
predicts the probability of each sound event type at each frame,
and these are aggregated across time using the max pooling function
to get clip-level probabilities of the sound events.

We used the stochastic gradient descent (SGD) algorithm to minimize
the cross entropy loss averaged over clips and sound event types.
We used a batch size of 95~clips.
We applied a Nesterov momentum of 0.8, and a gradient clipping limit of $10^{-3}$.
The learning rate was initialized to 0.1, and was decayed by a factor of 0.8
when the validation loss did not reduce for 3 consecutive epochs.

We implemented the network using the Keras toolkit~\cite{chollet2015keras},
and trained it on a single GeForce GTX 1080 Ti GPU.
Each epoch of training took 20~minutes, and the model would converge within 6~hours.

\subsection{Class-Specific Thresholding}

The clip-level probabilities predicted by the network must be thresholded
to generate binary predictions for evaluation.
Due to the imbalance between the sound event classes,
the distribution of predicted probabilities may vary drastically across them,
and we found it critical to tune the threshold for each class individually.
The primary evaluation metric used in the DCASE challenge is the $F_1$ score,
micro-averaged across all sound event types.
We devised an iterative procedure to tune class-specific thresholds
to optimize the micro-average $F_1$:
(1) tune the threshold of each class to maximize the class-wise $F_1$;
(2) repeatedly pick a random class and re-tune its threshold to optimize the micro-average $F_1$, until no improvements could be made.
We tuned the thresholds on the validation data after each epoch of training,
and picked the model reaching the highest $F_1$ on the validation data as the final model.
The thresholds obtained on the validation data were directly applied to the test and evaluation data.

\subsection{Fusion of Audio and Visual Information}

We used the procedure described in Section~\ref{sec:visual}
to estimate the probability of the sound events for each clip
using the visual information.
Each 10-second clip consists of 240 video frames;
we selected 4 key frames among them.
Object recognition and knowledge mapping produces
the probability of each of the 17 types of sound events.

To verify the quality of the video-only branch,
we calculated the $F_1$ score of each sound event type
using its predictions on the validation data.
The video predictions outperformed the audio-only baseline system
on seven sound events: ambulance, bicycle, bus, car passing by, fire truck, skateboard and truck.

We combined the outcome of the audio and video branches using different weights:
for the seven sound events where the video branch performed better,
we assigned a weight of 0.8 to the video branch and 0.2 to the audio branch;
for the remaining sound events, we gave a weight of 0.2 to the video branch and 0.8 to the audio branch.
These weights were not tuned to the fullest;
even better performance may be expected if they were tuned more carefully.

\subsection{Results and Analysis}

Table~\ref{table:dcase-performance} lists the micro-average $F_1$ score
on the development, test and evaluation set
of the baseline audio-only system and the multimodal system.
With the help of the visual information, we achieved a remarkable improvement of
10.5\%, 6.0\% and 5.3\% on the three sets, respectively.
The final $F_1$ score on the test set, 55.9\%,
is the best performance known so far for the audio tagging task of the DCASE 2017 challenge~\cite{Badlani2017}.

\begin{figure}
\centering
\begin{tikzpicture}
\begin{axis}[
    width=1.08\linewidth, height=6.0cm,
    xtick={1.0,...,17.0,18.0},
    xticklabels={
        CarPassingBy(\textbf{V}),
        Bus(\textbf{V}),
        Bicycle(\textbf{V}),
        Truck(\textbf{V}),
        ReverseBeeps,
        Ambulance(\textbf{V}),
        AirHorn,
        CarAlarm,
        Motorcycle,
        FireEngine(\textbf{V}),
        PoliceCar,
        TrainHorn,
        Car,
        Train,
        Skateboard(\textbf{V}),
        Screaming,
        CivilSiren},
    xmajorgrids=true,
    x tick label style={rotate=90,anchor=east},
    grid=major,
    ybar=1*\pgflinewidth,
    bar width=4pt,
    enlarge x limits=0.03,
    legend style={at={(0.5,0.95)}, anchor=south,legend columns=-1}
]

\addplot[
    fill=bblue,
    draw=black,
    point meta=y,
    every node near coord/.style={inner ysep=3pt},
    error bars/.cd,
    y dir=both,
    y explicit
]
coordinates{
    (1,0.000)
    (2,0.321)
    (3,0.387)
    (4,0.387)
    (5,0.429)
    (6,0.500)
    (7,0.571)
    (8,0.579)
    (9,0.579)
    (10,0.585)
    (11,0.588)
    (12,0.712)
    (13,0.721)
    (14,0.726)
    (15,0.737)
    (16,0.747)
    (17,0.797)
};

\addplot[
    fill=ggreen,
    draw=black,
    point meta=y,
    every node near coord/.style={inner ysep=3pt},
    error bars/.cd,
    y dir=both,
    y explicit
]
coordinates{
    (1,0.286)
    (2,0.518)
    (3,0.517)
    (4,0.541)
    (5,0.518)
    (6,0.607)
    (7,0.557)
    (8,0.624)
    (9,0.623)
    (10,0.585)
    (11,0.503)
    (12,0.762)
    (13,0.771)
    (14,0.772)
    (15,0.777)
    (16,0.858)
    (17,0.837)
};

\legend{Audio Only, Audio + Video}
\end{axis}
\end{tikzpicture}
\caption{Class-wise $F_1$ scores of the audio-only system vs. the audio-visual reranked system on the test data.
(\textbf{V}) indicates sound events for which the visual information receives a higher weight in the reranking.} \label{fig:dcase-performance}
\end{figure}
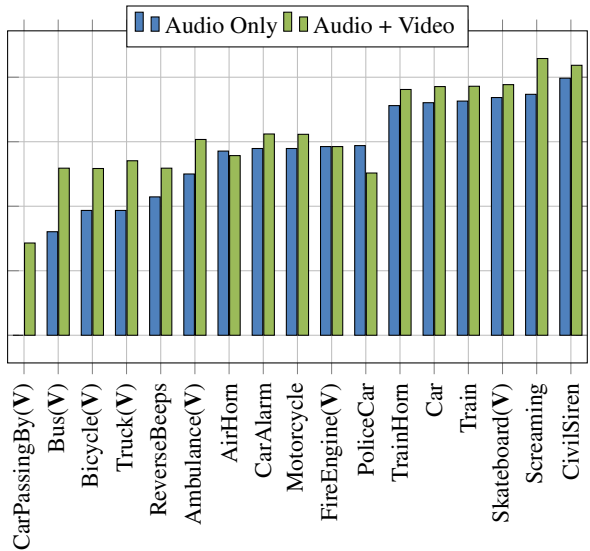

The contribution of visual information to the test $F_1$ score of each sound event type
is shown in Fig.~\ref{fig:dcase-performance}.
The fusion improved the performance of 14 out of the 17 sound event types.
The most significant improvement came from the ``car passing by'' class.
This class is confusable with the ``car'' class and contains very few positive examples,
and was totally missed by the baseline system.
Most of the sound event types related to types of vehicles
also saw an improvement, because the video tracks clearly show the type of vehicle involved.
The ``bicycle'' class benefited from visual information because
may clips show people talking about bicycles,
in which bicycles are visible in the video track but not audible in the audio track.
The analysis of this result reveals some problems of the data annotation.

We would like to emphasize that the entire multimodel system
consumed very few computational resources.
The CRNN in the audio branch took 6 hours to train on a GPU;
the video branch, which only used pretrained models,
took less than 30 minutes to run on the validation, test and evaluation data.

\section{Conclusion \& Future Work}
\label{sec:conclusion}

In this work, we have proposed a multimodal framework for audio tagging,
which combines information from both the audio and the video tracks of video clips
to predict the sound events in the audio track.
The system outperforms a strong baseline system
on the audio tagging task of the recent DCASE 2017 challenge,
boosting the test $F_1$ score by 5.3\%.
This framework is also light-weight compared to other models that use visual information,
because it makes extensive use of pretrained state-of-the-art deep learning models
and avoids training them from scratch. In the future, we would delve more into understanding the correspondence between audio and video and explore the optimal multi-modal system that balances the performance and training resource trade-off.

\begin{table}
  \centering
  \begin{tabular}{|c|c|c|c|}
  \hline
    \textbf{System} & \textbf{Development} & \textbf{Test} & \textbf{Evaluation} \\
    \hline
    \textbf{Audio Only} & 50.1 & 54.9 & 50.6 \\
    \hline
    \textbf{Audio + Video} & 60.6 & 60.9 & 55.9 \\
    \hline
  \end{tabular}
  \caption{Micro-average $F_1$ score (in percents) of the audio-only system and the audio-visual reranked system on development, test and evaluation data.}
  \label{table:dcase-performance}
\end{table}

\bibliographystyle{IEEEbib}
\bibliography{Master}

\end{document}